\documentclass{aa}  

\usepackage{graphicx}
\usepackage{txfonts}
\usepackage{hyperref}
\usepackage{amsmath}	
\usepackage[version=4]{mhchem}
\usepackage[normalem]{ulem}
\usepackage{xcolor}
\usepackage{cleveref}

\begin{document}

   \title{New routes for PN destruction and formation in the ISM via neutral-neutral gas-phase reactions  and an extended database for reactions involving phosphorus}


   \author{Mateus X. Silva
          \inst{1}
          \and
          Edgar Mendoza\inst{2}
          \and
          Fábio S. L. Ferreira\inst{3}
          \and
          Alexandre C. R. Gomes\inst{3,5}
          \and
          Miguel Carvajal\inst{2,4}
          \and
          Jing Li\inst{5}
          \and
          António J. C. Varandas\inst{1,5,6}
          \and
          Breno R. L. Galvão\inst{3}
          }

   \institute{Departamento de Física, Universidade Federal do Espírito Santo, Av. Fernando Ferrari 514, Vitória, ES, 29075-910, Brazil
        \and
             Dept. Ciencias Integradas, Facultad de Ciencias Experimentales, Centro de Estudios Avanzados en Física, Matemática y Computación, Unidad Asociada GIFMAN, CSIC-UHU, Universidad de Huelva, Spain\\
             \email{edgar.mendoza@dci.uhu.es}
        \and
            Departamento de Química, Centro Federal de Educação Tecnológica de Minas Gerais, Av. Amazonas 5253, Belo Horizonte, MG, 30421-169, Brazil            
        \and
            Instituto Universitario "Carlos I" de Física Teórica y Computacional, Universidad de Granada, Spain
        \and
            School of Physics and Physical Engineering Qufu Normal University, Qufu, Shandong 273165, China
        \and
            Coimbra Chemistry Centre and Chemistry Department, University of Coimbra, 3004-535 Coimbra, Portugal
             }

   \date{Received November XX, XXXX; accepted XXX XX, XXXX}

 
  \abstract
   {Phosphorus plays an essential role in the chemistry of living organisms, being present in several fundamental biomolecules. The investigation of chemical reactions taking place in different astronomical environments involving phosphorus-containing molecules is essential for understanding how these species are  produced and destroyed. Ultimately, it can help unravel the pathways that lead to important prebiotic molecules.}
   {Phosphorus monoxide (PO) and phosphorus nitride (PN) are key reservoirs of phosphorus in the Interstellar Medium (ISM). Understanding their reaction mechanisms and accurately predicting rate coefficients are crucial for modeling phosphorus chemistry in space. This work presents a computational study of the CPN system to identify viable reaction pathways involving atom-diatom collisions and to explore a potential destruction route for PN in the ISM. We also evaluate the role of several neutral-neutral reactions involving PO and PN in chemical models simulating interstellar environments.}
   {In this work we explore the potential energy landscape of the C($\mathrm{^3P}$) + PN($^1\Sigma^+$), N($\mathrm{^4S}$) + CP($^2\Sigma^+$) and P($\mathrm{^4S}$) + CN($^2\Sigma^+$) reactions by performing high-accuracy \textit{ab initio} calculations and provide their rate coefficients over a wide range of temperatures. The temperature-dependent rate coefficients were fitted to the modified Arrhenius equation: $k(T)=\alpha(T/300)^{\beta}\mathrm{exp}(-\gamma/T)$. An updated chemical network for P-bearing species was used to model the time-dependent abundances and reaction contributions of P, PO, PN, and PH during the chemical evolution of diffuse/translucent and dense clouds.}
   {The only neutral-neutral reaction capable of destroying PN without an activation energy seems to be the PN+C one. We have also shown that reactions between CP and N can yield CN and PN barrierless. Chemical models indicate that PO is a crucial species driving the gas-phase formation of PN. Typically, PO/PN ratios exceed 1, though their chemistry is  influenced by photon- and cosmic-ray-induced processes. Over time in simulated dense clouds, neutral-neutral reactions such as PO + N, PH + N,  P + OH, and PH + O play a significant role in determining the relative abundances of PO and PN.}
   {}

\titlerunning{Destruction of PN through the C+PN reaction}

   \maketitle
%

\section{Introduction}

Phosphorus is present in several molecules that are essential for life as we know it, although it is still a mystery how it became biologically available on Earth, possibly from an inorganic mineral or even extraterrestrial sources \citep{PAS05:515,SCH06:1743}. When combined with four oxygen atoms, it yields the phosphate ion ($\mathrm{PO_4^{3-}}$), which is highly soluble in water and whose salts are plentiful on Earth and were readily available throughout the development of life \citep{HUN12:2513}.

There is now biochemical evidence which indicates that phosphate may indeed have played an important role in the chemical origins of life. Several feasible environments capable of producing phosphates are currently conceivable, providing numerous potential solutions to the issue of phosphate availability \citep{FER17:31}. Phosphorus, as well, can form P-N (phosphoramidate), P-S (phosphorothioate) and P-C (phosphonate) linkages, all of which are present in nature. For instance, phosphorylation of amino acids such as histidine, lysine and arginine to generate high energy phosphoramidate bonds may have been crucial under prebiotic circumstances \citep{HUN12:2513}.

The prebiotic chemistry (including  P-bearing molecules) on our planet may have been enriched by the bombardment of extraterrestrial bodies such as comets and asteroids \citep{PAS05:515,ALT16:e1600285,MAR23:47}. For this reason, the investigation of chemical processes taking place on different astronomical environments such as molecular clouds and star forming-regions is of great importance to our understanding of the primordial chemistry of phosphorus on Earth.

The detection of most P-bearing species in space has been restricted so far to the circumstellar envelopes of evolved stars \citep{GUE90:L9,AGU07:L91,HAL08:L101,MIL08:618,AGU14:A45,AGU14:L27} in our Galaxy and planetary atmospheres \citep{BRE75:L55,LAR77:972}. However, special attention has been given to phosphorus monoxide (PO) and phosphorus nitride (PN), which were the only ones also detected towards star-forming regions \citep{FON16:L30,RIV16:161,MIN18:L39,BER19:L36,FON19:4530,RIV20:1180,BER21:55}, in addition to the interstellar medium \citep{TUR87:L75,ZIU87:L81}, circumstellar envelopes of evolved stars \citep{TEN07:L29,MIL08:618,ZIU18:169}, protostellar shocks \citep{YAM11:L37,LEF16:3937} and Galactic Centre molecular clouds \citep{RIV18:L30}.

It has already been discussed that PO and PN may constitute the main reservoir of phosphorus in the gas phase in the ISM \citep{THO84:139,TEN07:L29,BEC13:A132,LEF16:3937}. In low-mass star-forming regions, PN was the first P-bearing molecule to be detected \citep{YAM11:L37},
while PO was first detected five years later towards the same solar-type star-forming region (L1157-B1) \citep{LEF16:3937}. The abundances of these species in such environments seem to be comparable, with PO being systematically more abundant than PN \citep{RIV16:161,LEF16:3937,RIV18:L30,BER19:L36}. However, the small number of detection makes it difficult to precisely elucidate the mechanisms of formation of these molecules.

More recently, the presence of PO and PN in the gas phase has also been reported for the first time at the edge of our galaxy \citep{KOE23:292}. The characteristic rotational lines of these P-bearing species were observed in the dense cloud WB89-621 (located 22.6 kpc from the Galactic Centre) and their abundances are comparable to values near the Solar System \citep{BER21:55,KOE23:292}. Specifically, the phosphorus nitride molecule has been detected for the first time in an extra-galactic environment towards two giant molecular clouds of NGC 253 \citep{HAA22:A158}. The authors report a derived abundance ratio between PN and the shock-tracer SiO that follows the same trend already found towards Galactic sources \citep{RIV18:L30,HAA22:A158}.

Noticeably, the [PO]/[PN] ratio has consistently been found to be > 1, independently of the source, with values up to $\sim6$ \citep{HAA22:A158,MAR23:47}. However, the available astrochemical models most frequently fail to reproduce such trend, usually predicting [PO]/[PN] ratios < 1 for various different physical conditions \citep{JIM18:128J,CHA20:A54,SIL21:119,MAR23:47}. Various factors could contribute to such disparities, including the inadequacy of the phosphorus chemical network and the considerable uncertainties associated with reaction rate coefficients \citep{MAR23:47}. Currently, there is a compelling evidence indicating that PO emerges in the gaseous phase prior to PN and then vanishes, whereas PN lingers on for a certain time \citep{GOM23:5991}. The precise mechanisms accountable for their formation and decomposition are yet to be better delineated and their study may solve the inconsistency between astrochemical models and observations.

A computational approach can help in understanding the chemistry of phosphorus in the gas phase since experimental studies involving these P-bearing compounds are scarce in the literature. \citet{Garcia2021} obtained rate coefficients and branching ratios for the reactions P + OH $\rightarrow$ PO + H and P + $\mathrm{H_2O}$ $\rightarrow$ PO + $\mathrm{H_2}$ using \textit{ab initio} transition state theory (TST) and employing high-level quantum chemical calculations. The authors conclude that the formation of PO, in the gas phase, from OH and P in their ground state is highly favourable, but should not occur if P is in its first excited electronic state. As for the reaction between P and $\mathrm{H_2O}$, the predicted trend is the opposite. More recently, the P($\mathrm{^4S}$) + $\mathrm{O_2}$($^3\Sigma^-$) and P($\mathrm{^4S}$) + $\mathrm{O_2}$($^1\Delta$) reactions were also investigated by \citet{GOM22:259} and \citet{CON24:142} from a theoretical point of view, and the mechanisms towards PO formation could be elucidated. It has been concluded that multireference methods are needed to properly describe the system and that the P($\mathrm{^4S}$) + $\mathrm{O_2}$($^3\Sigma^-$) reaction dominates the formation of PO when OH is not sufficiently available (for instance, in regions where shock speeds are below $40$\,km\,s$^{-1}$). Structural, molecular and spectroscopic properties of the HPO system were studied using the coupled cluster method by \citet{FRA03:303} and \citet{PUZ06:238}, and isomerisation and dissociation energies were obtained, showing that the two processes may compete with each other.

\citet{PLA21:10} estimated a series of rate coefficients for several reactions of the $\mathrm{H_xPO_y}$ system from electronic structure calculations and subsequent application of statistical rate theory. In the following year, \citet{DOU22:99} gathered an extensive network of reactions involving phosphorus with estimates for their rate coefficients obtained from electronic structure calculations at the Complete Basis Set (CBS-QB3) level \citep{MON00:6532}.

Our research group has also approached the NPO system using high-level electronic structure methods \citep{SOU21:2} and quantitatively confirmed that the depletion of PO by N atoms occurs fast, with a branching ratio favourable to the O + PN channel relative to P + NO. The HPN system was also addressed \citep{GOM23:e011} using \textit{ab initio} calculations and the variable reaction coordinate transition state theory (VRC-TST) and it was predicted that the reactions P($\mathrm{^4S}$) + NH($^3\Sigma^-$) and N($\mathrm{^4S}$) + PH($^3\Sigma^-$) occur without an activation barrier and are relevant for PN formation in the interstellar medium.

Astrochemical models, however, appear to lack important PN destruction pathways \citep{MAR23:47}. This may be one reason for the failure to correctly predict the [PO]/[PN] ratio. Previous studies from our research group indicate that PN destruction by H atoms should not occur \citep{GOM23:e011}. However, PN destruction by N atoms may occur in high temperature environments, such as in shock regions \citep{GOM23:5991}. \citet{JIM18:128J} drew attention to the fact that the reactions N + CP $\rightarrow$ PN + C and P + CN $\rightarrow$ PN + C, proposed by \citet{AGU07:L91}, have very uncertain rate coefficients and that the results of different models are little affected by the inclusion of these reactions. In the work of \citet{MAR23:47}, the parameters associated with the analogous reaction N + CN $\rightarrow$ $\mathrm{N_2}$ + C were used in the rate coefficients of the N + CP and P + CN reactions, assuming that the similarity between N and P should lead to similar chemical behaviour.

Therefore, in the present work, we focus on the computational study of the CPN system using the highest-level electronic structure methods available combined with TST. The main goals are to elucidate the viable reaction pathways involving simple atom-diatom collisions within this system (C($\mathrm{^3P}$) + PN($^1\Sigma^+$), N($\mathrm{^4S}$) + CP($^2\Sigma^+$) and P($\mathrm{^4S}$) + CN($^2\Sigma^+$)), providing the respective rate coefficients and evaluating a possible route for the destruction of PN in the interstellar medium. We summarize other relevant reactions from the literature, and include them in the KIDA \citep{KIDA} network to investigate the most important reactions that drive the abundances of PO and PN in the ISM.

\section{Methodology}

\subsection{Electronic structure calculations}
\label{sec:ee-calc}

All electronic structure calculations reported here were performed using the MOLPRO \citep{MOLPRO_brief} package. For the initial exploration of the potential energy surface (PES) of the PCN system, a density functional theory (DFT) \citep{DFT} approach was employed using the aug-cc-pV(T+d)Z \citep{DUN89:1007,KEN92:6796} basis set (henceforth referred to as AV(T+d)Z) with the M06-2X \citep{ZHA08:215} exchange and correlation functional. At this stage, stationary points were found and vibrational analysis was performed to confirm whether the obtained structures corresponded to energy minima or to first order saddle points on the PES. Intrinsic reaction coordinate (IRC) calculations were performed for all transition states (TSs) found in order to ensure their connection to the correct energy minima structures. All the obtained energies were zero-point energy (ZPE) corrected. In order to improve the accuracy of the energy values, further single-point energy calculations at the explicitly correlated coupled cluster (CCSD(T)-F12/cc-pVTZ-F12) \citep{ADL07:221106,KNI09:054104} level were performed over the geometries optimized at the DFT level and we name this methodology CCSD(T)-F12/VTZ-F12//M06-2X/AV(T+d)Z + ZPE(M06-2X/AV(T+d)Z).  We have not employed scaling factors in the vibrational frequencies.

In a second stage, in order to account for a possible multiconfigurational character of the PCN system, all the previously obtained results were fully re-optimised within the \textit{ab initio} full valence complete active space self-consistent field (CASSCF) \citep{SZA12:108} level with AV(T+d)Z basis set. In this approach, the 1s orbitals of C and N, and the 1s, 2s and 2p orbitals of P were considered as core orbitals, leading to an active space of 12 orbitals with 14 electrons. Vibrational analysis was also carried out and ZPEs were obtained for all stationary points on the CASSCF PES. This time, the accuracy of the energy values was enhanced by performing single-point energy calculations over all structures with the explicitly correlated multireference configuration interaction (MRCI-F12) \citep{SZA12:108,SHI11:034113} method, including the relaxed-reference Davidson correction to approximate quadruple excitations and also using the AV(T+d)Z basis set. The results reported for this methodology are referred here as MRCI-F12/AV(T+d)Z//CAS/AV(T+d)Z + ZPE(CAS/AV(T+d)Z).

The methodologies employed here could also provide the magnitude of the dominant configuration in the CI wavefunctions ($\mathrm{C_0^2}$) and the $\mathrm{T_1}$ diagnostic of all stationary points, which allows the assessment of the multireference character (MR) of the system \citep{LEE89:199,LEI00:431,LEE03:362}. Lastly, the Wxmacmolplt and Avogadro software were used for graphical visualisation and representation of the molecular geometries \citep{BOD98:133,HAN12:17}.

\subsection{Rate coefficients}
\label{sec:coeff}

In order to assess the effects of the reactions studied in this work on the relative abundances of the involved species, their rate coefficients must be calculated so that the chemical processes taking place in an interstellar environment can be simulated. We have obtained rigid rotor/harmonic oscillator rate coefficients using standard TST and including the Eckart tunnelling correction \citep{ECK30:1303} employing the MESS software package \citep{GEO13:12146}. Phase space theory (PST) \citep{PEC65:3281,CHE86:2615}, as implemented in MESS, is used for the barrierless capture steps. For such cases, we have fitted the MRCI-F12/AV(T+d)Z energies obtained along the minimum energy path as a function of the internuclear separation ($r$) to the $-C_6/r^6$ function to obtain $C_6$, which is then used in the PST calculation.

\subsection{Astrochemical modelling}
\label{sec:gas-grain}

We utilized the Nautilus gas-grain code to simulate chemical abundances under interstellar conditions, focusing on the gas-phase and grain-surface processes described in the three-phase model by \citet{Ruaud2016}. The gas-phase chemical network is based on the comprehensive kida.uva.2014 database \citep{Wakelam2015}, which compiles extensive reaction pathways and kinetic data. Although the kida.uva.2024 update has introduced new species and reaction channels, it does not modify phosphorus chemistry \citep{Wakelam2024}. In this work, we expanded the chemical network within Nautilus to include updated reaction pathways for phosphorus-containing molecules. This enabled the prediction of abundances for key astrochemical species such as PO and PN, whose formation mechanisms in extreme environments remain poorly understood. The study also assessed the significance of individual reaction pathways in phosphorus chemistry. Cosmic ray and X-ray ionization rates, crucial drivers of interstellar chemical reactions, were systematically varied in the simulations. The default cosmic ray ionization rate ($\zeta_{\rm CR}$) is typically set to $1.3 \times 10^{-17}$ s$^{-1}$ \citep{Dalgarno2006}, but can reach $10^{-15}$ s$^{-1}$ in regions like the central molecular zone of Sgr B2 \citep{Goto2014}. Similarly, X-ray ionization rates ($\zeta_{\rm X}$) influence the chemistry of small neutrals and molecular ions, particularly in protostellar envelopes \citep{Stauber2005, Notsu2021}. We tested $\zeta_{\rm CR}$ values from $1.3 \times 10^{-17}$ to $1.3 \times 10^{-16}$ s$^{-1}$ and $\zeta_{\rm X}$ values from $1 \times 10^{-18}$ to $1 \times 10^{-17}$ s$^{-1}$ to evaluate their effects on phosphorus chemistry.

Simulations were conducted to examine phosphorus chemistry under two distinct astrophysical conditions: diffuse/translucent clouds and dense clouds \citep{Corbi2018,Phan2018,Wolfire2022}. In the diffuse/translucent  model, the H$_2$ gas density was set to 10$^3$ cm$^{-3}$, with a temperature of 70~K and visual extinction ($A_V$) less than 5~mag. The simulation covered a timescale of 1 Myr, during which cosmic ray and X-ray ionization rates were varied to assess their impact on chemical processes. For the dense molecular cloud model, the gas density was increased to 10$^4$ cm$^{-3}$, while the temperature was lowered to 30 K and $A_V$ was fixed at 10~mag. Simulations extended from 1 $\times 10^6$ yr up to 1 $\times 10^8$ yr to explore long-term chemical evolution. Ionization rates were held constant at $\zeta_{\rm CR}$ = $1.3 \times 10^{-17}$ s$^{-1}$ and $\zeta_{\rm X}$ = 1 $\times$ 10$^{-18}$ s$^{-1}$. In both models, abundances were calculated as functions of time. The non-linear nature of the chemical network and the dynamic interplay of reactions prevented the system from reaching full equilibrium, influencing the steady-state abundances \citep{Roueff2020,Dufour2023}. In the dense cloud model, extended simulations provided insights into the stability and evolutionary patterns of phosphorus-bearing species. Individual reaction contributions were evaluated to quantify their roles in shaping chemical pathways and the long-term persistence of these molecules in interstellar environments.

\section{Results}

\subsection{Reaction products}
\label{sec:prod}

As we expect the reactants (C($\mathrm{^3P}$) + PN($^1\Sigma^+$)) to be mainly available in their electronic ground state, we have only considered electronic states with triplet spin multiplicities. The $\mathrm{C}_s$ symmetry of the system was taken into consideration and the first electronic level of each irreducible representation ($A'$ and $A''$) was calculated.  
The gas-phase chemical reactions investigated in this work, with the respective estimated energies at MRCI-F12/AV(T+d)Z//CAS/AV(T+d)Z + ZPE(CAS/AV(T+d)Z) level, are listed below:

\begin{align}
\label{eq:c+pn1} \ce{C(^3P) + PN(^1\Sigma^+)} &\ce{-> P(^4S) + CN(^2\Sigma^+)} &-140\,\mathrm{kJ\,mol^{-1}}\\
\label{eq:c+pn2} &\ce{ -> P(^4S) + CN(^2\Pi)} &-33\,\mathrm{kJ\,mol^{-1}}\\
\label{eq:c+pn3} &\ce{ -> N(^4S) + CP(^2\Sigma^+)} &83\,\mathrm{kJ\,mol^{-1}}\\*[5mm]
\label{eq:n+cp} \ce{N(^4S) + CP(^2\Sigma^+)} &\ce{-> P(^4S) + CN(^2\Sigma^+)} &-223\,\mathrm{kJ\,mol^{-1}}
\end{align}

The only exoergic products obtained from the C + PN reactants are P + CN. Note that the formation of excited state cyano radical CN$(^2\Pi)$ is also exoergic. The cyano radical has been widely observed in several astronomical environments \citep{MCK40:187,ADA41:11,JEF70:L87,FED84:219,HEN88:L23,FRA05:1243,HAN15:A131,PAR21:A77} and one should expect it to be often involved in high energy chemistry. Indeed, from the reactions studied here, CN is not expected to be destroyed by collisions with P (no exoergic products available).

Our calculations also indicate that carbon monophosphide (CP), which has already been detected in space \citep{GUE90:L9}, may be promptly destroyed by collisions with neutral N atoms in the highly exoergic N+CP\ce{->} P+CN reaction (Eq.~\ref{eq:n+cp}). Another possible pathway for CP destruction by N atoms is N + CP\ce{->} C + PN (Eq.\,$-$~\ref{eq:c+pn3}, the reverse of Eq.~\ref{eq:c+pn3}) with an energy release of $-83$ kJ\,mol$^{-1}$. Our computational results on this reaction corroborates the conclusion of \citet{CHA20:A54} that this reaction may be an important route of PN formation. In addition, \citet{CHA20:A54} also predict that the main formation pathway for PN in the early stage of evolution of diffuse and translucent clouds is \ce{P + CN -> C + PN}, while according to our calculations this reaction (Eq.\,$-$~\ref{eq:c+pn1}, the reverse of Eq.~\ref{eq:c+pn1}) is estimated to be significantly endoergic. Therefore, it should not contribute to PN formation in cold environments such as the ones modelled in the work of \citet{CHA20:A54}.

In fact, \citet{CHA20:A54} included the \ce{N + CP -> C + PN} and \ce{P + CN -> C + PN} reactions in their chemical network of phosphorus as proposed by \citet{AGU07:L91}, which employed a rate coefficient equal to that of the nitrogen analogous reaction (\ce{N + CN -> C + N_2}). The value of $k=3\times 10^{-10}$\,cm$^{3}$\,s$^{-1}$ was recommended over the range 300-2500 K and, at low temperatures, the measurements are reported to scatter over four orders of magnitude \citep{BAU92:411}. Moreover, the \ce{N + CN -> C + N_2} reaction has also been studied computationally \citep{MOS01:4156} and is predicted to be highly exothermic (with an energy release of $-205$ kJ\,mol$^{-1}$), presenting an energy landscape that is very different from that of the true chemical process investigated here. The use of the nitrogen analogous reaction is thus inadequate to properly describe the chemical evolution of these P-bearing species in cold regions of the ISM.

\subsection{Potential energy surface and reaction mechanisms}
\label{sec:mec}

Along with the thermodynamic analysis presented in the previous section, an investigation of the kinetic viability of the reactions is necessary. The existence of small potential energy barriers, for instance, can turn the formation of thermodynamically favourable products negligible in cold parts of the ISM.

\begin{figure*}
\includegraphics[width=1.0\textwidth]{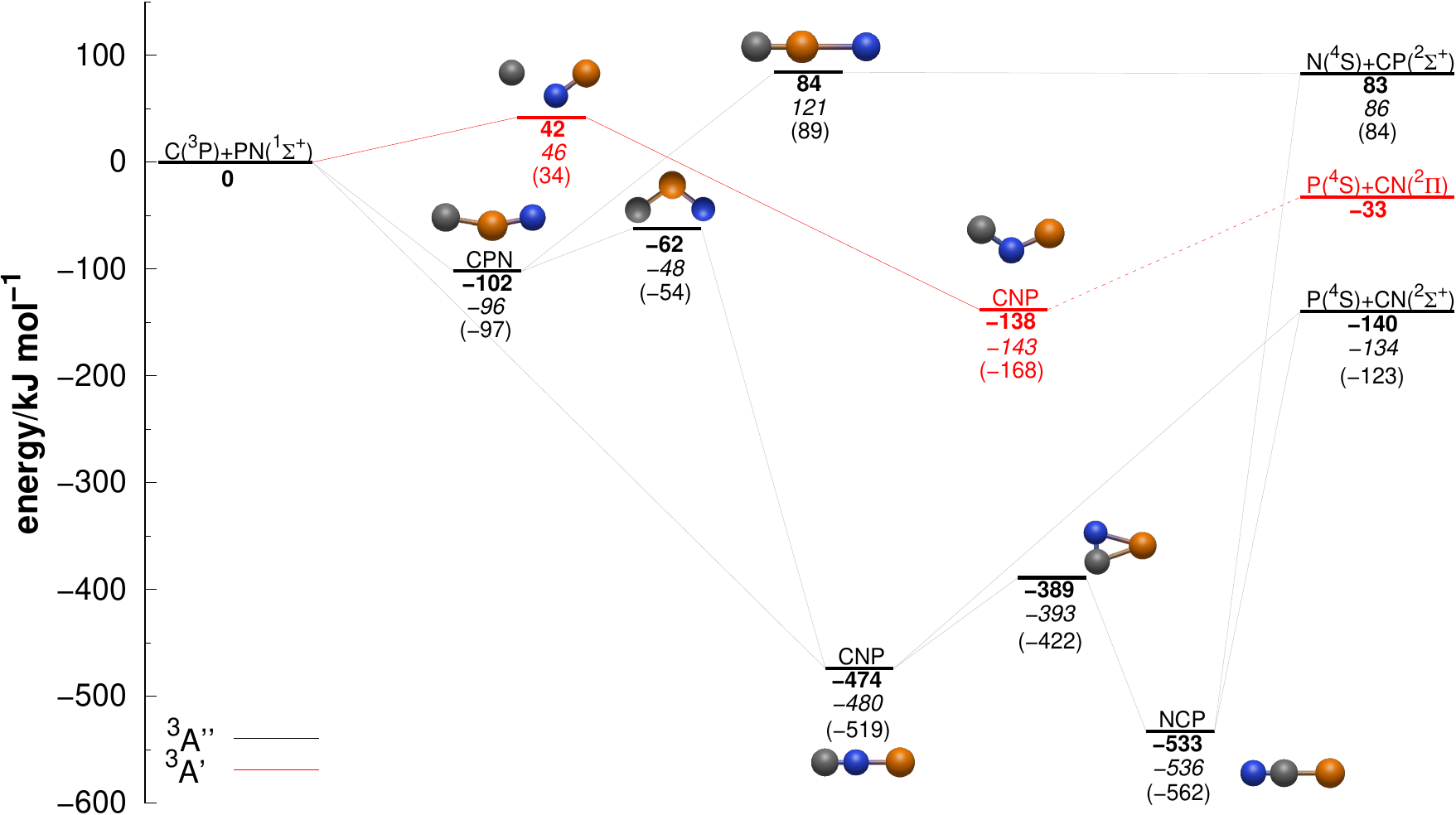}
    \caption{Potential energy diagram for the $^3A'$ (red) and $^3A''$ (black) electronic states of the CPN system. Energies are given in kJ\,mol$^{-1}$ relative to the C+PN limit and are ZPE corrected. The MRCI-F12 energies are given in boldface, CCSD(T)-F12 results are given in italics and M06-2X ones are given under parenthesis. Atoms are coloured as: carbon (grey); phosphorus (orange); nitrogen (blue).}
    \label{fig:pes}
\end{figure*}

To elucidate the possible reaction pathways of the CPN system, we have explored its triplet $A'$ and $A''$ potential energy surfaces and obtained four relevant potential energy minima and four TSs. Table~\ref{tab:struct-param} gathers their geometries and energies, while Fig.~\ref{fig:pes} shows a potential energy diagram summarising our results and the connections between the stationary structures. The Cartesian coordinates and frequencies of all structures can be found in the supplementary information.

Starting with the C+PN collision, the ground state for both product channels (P+CN and N+CP) can only be achieved adiabatically in the $^3A''$ state (see Fig.~\ref{fig:pes}). The CN radical in its first excited electronic state can also be obtained adiabatically, but on the $^3A'$ surface. Despite being an exoergic process, it involves surpassing an energy barrier of 42 kJ\,mol$^{-1}$ in the entrance channel for the C($\mathrm{^3P}$) + PN($^1\Sigma^+$) reaction, which should inhibit the formation of excited  CN(A$^2\Pi$) at low temperatures.

\begin{table}
	\centering
	\caption{Properties of the asymptotic limits and stationary points of $^3A'$ and $^3A''$ PESs$^a$.}
	\label{tab:struct-param}
	\begin{tabular}{lcccr} 
		\hline
		   & $R_{PN}$ & $R_{PC}$ & $R_{CN}$ & E$^b$\\
		\hline
		\multicolumn{1}{c}{ {$^3A'$ }}\\*[1mm]
		CNP & 1.59 & 2.59 & 1.40 & -138\\
		$\mathrm{TS_{C+PN\rightarrow CNP}}$ & 1.54 & 3.05 & 2.04 & 42\\*[2mm]
            \multicolumn{1}{c}{ {$^3A''$ }}\\*[1mm]
            CPN & 1.54 & 1.77 & 3.26 & -102\\
            CNP & 1.70 & 2.89 & 1.19 & -474\\
            NCP & 2.95 & 1.77 & 1.18 & -533\\
            $\mathrm{TS_{CPN\rightarrow N+CP}}$ & 2.50 & 1.59 & 4.10 & 84\\
            $\mathrm{TS_{CPN\rightarrow CNP}}$ & 1.63 & 1.75 & 2.72 & -62\\
            $\mathrm{TS_{CNP\rightarrow NCP}}$ & 2.01 & 1.99 & 1.21 & -389\\*[2mm]
            \multicolumn{5}{l}{Asymptotic limits }\\*[1mm]
            C($\mathrm{^3P}$) + PN($^1\Sigma^+$) & 1.51 & - & - & 0\\
            N($\mathrm{^4S}$) + CP($^2\Sigma^+$)) & - & 1.58 & - & 83\\
            P($\mathrm{^4S}$) + CN($^2\Pi$)) & - & - & 1.25 & -33\\
            P($\mathrm{^4S}$) + CN($^2\Sigma^+$)) & - & - & 1.18 & -140\\
		\hline
	\end{tabular} \\
    \footnotesize{$^a$ Interatomic distances are in given in \AA \, and energies in kJ\,mol$^{-1}$ relative to the C+PN limit.}\\
    \footnotesize{$^b$At the MRCI-F12/AV(T+d)Z level including ZPE correction calculated at the CASSCF/AV(T+d)Z level.}
\end{table}

It can be seen from Fig.~\ref{fig:pes} that the relative energies obtained at CCSD(T)-F12 and MRCI-F12 levels are in good agreement, deviating from each other by a maximum of $6$ kJ\,mol$^{-1}$, except for the $\mathrm{TS_{CPN\rightarrow CNP}}$ and $\mathrm{TS_{CPN\rightarrow N+CP}}$ cases. This may be attributed to a high multiconfigurational character of these structures (the values for the $\mathrm{T_1}$ diagnostic is 0.056 and 0.080, respectively). The M06-2X results agree qualitatively with the MRCI-F12 ones, with relative energy deviations mostly within $10$\%. The largest deviation observed in the DFT approach was for the CNP ($^3A'$) minimum, reaching about $22$\% (an absolute deviation of $30$ kJ\,mol$^{-1}$). From now on the discussion will be based on the more trustworthy MRCI-F12/AV(T+d)Z//CAS/AV(T+d)Z + ZPE(CAS/AV(T+d)Z) results.

We first address the possible routes of PN destruction by a carbon atom. If the incoming C atom attacks the phosphorus end of the PN molecule, one can notice in Fig.~\ref{fig:pes} that this can lead to a bent CPN intermediate through a barrierless path on the $^3A''$ PES. Subsequently, this minimum can either isomerize to a linear CNP intermediate or dissociate to N($\mathrm{^4S}$) + CP($^2\Sigma^+$). Such abstraction mechanism \ce{C + PN -> CPN -> N + CP} requires, nevertheless, the surpassing of a $84$ kJ\,mol$^{-1}$ reaction barrier (relative to C + PN) that separates the CPN minimum from the products. Besides the high energy barrier, the products themselves lie $83$ kJ\,mol$^{-1}$ above the reactants. Therefore, this mechanism can be considered closed for typical ISM conditions. Furthermore, the isomerisation of the CPN minimum to the CNP one occurs through a conventional TS that lies $40$ kJ\,mol$^{-1}$ above CPN, but is submerged relatively to reactants. The CNP isomer can also undergo a second isomerisation to yield the linear NCP intermediate, which corresponds to the global minimum of this triatomic system. This second isomerisation also occurs through a conventional TS that is also submerged relatively to reactants. Direct dissociation from the CNP and NCP minima can lead to P($\mathrm{^4S}$) + CN($^2\Sigma^+$), without an exit barrier. The \ce{C + PN -> P($\mathrm{^4S}$) + CN($^2\Sigma^+$)} reaction is thus a viable PN destruction route  (exoergic and without an activation energy). It is open for whatever temperature, and may be important to model the abundance of PN in the ISM.

The P($\mathrm{^4S}$) + CN($^2\Sigma^+$) products can also be obtained through a simpler mechanism. For example, by the initial approach of the carbon atom to the diatomic PN via the N atom to form the CNP intermediate, which can then lose the P atom in an overall barrierless process (\ce{C + PN -> CNP -> P($\mathrm{^4S}$) + CN($^2\Sigma^+$)}).

In summary, the $^3A''$ PES allows for the possibility of several mechanisms of PN destruction by a carbon atom, all of which are listed below:

\begin{align}
\label{eq:mec1} &\ce{R -> CNP -> P(^4S) + CN(^2\Sigma^+)}\\
\label{eq:mec2} &\ce{R -> CNP -> NCP -> P(^4S) + CN(^2\Sigma^+)}\\
\label{eq:mec3} &\ce{R -> CPN -> CNP -> P(^4S) + CN(^2\Sigma^+)}\\
\label{eq:mec4} &\ce{R -> CPN -> CNP -> NCP -> P(^4S) + CN(^2\Sigma^+)}
\end{align}

\noindent where R means reactants (C($\mathrm{^3P}$) + PN($^1\Sigma^+$)).

When it comes to N + CP collisions, one can see in Fig.~\ref{fig:pes} that it has two exoergic product channels that may compete: C($\mathrm{^3P}$) + PN($^1\Sigma^+$) and P($\mathrm{^4S}$) + CN($^2\Sigma^+$). The abstraction of a carbon atom by the incoming nitrogen atom to produce P($\mathrm{^4S}$) + CN($^2\Sigma^+$) can easily take place through a direct mechanism that involves the formation of the global minimum of this triatomic system. This \ce{N + CP -> NCP -> P($\mathrm{^4S}$) + CN($^2\Sigma^+$)} pathway is barrierless and should be open regardless of the collision energy. The abstraction of the P atom by the incoming N atom to yield C + PN, in turn, presents a small energy barrier in the entrance channel. This barrier of just $1$ kJ\,mol$^{-1}$ associated with the \ce{N + CP -> CPN -> C + PN} reaction path lies below the accuracy of our quantum chemical calculations, but this reaction can also occur without an activation energy through the N+CP\ce{->}NCP \ce{->} CNP\ce{->} C+PN mechanism.

\begin{figure}
	\includegraphics[width=\columnwidth]{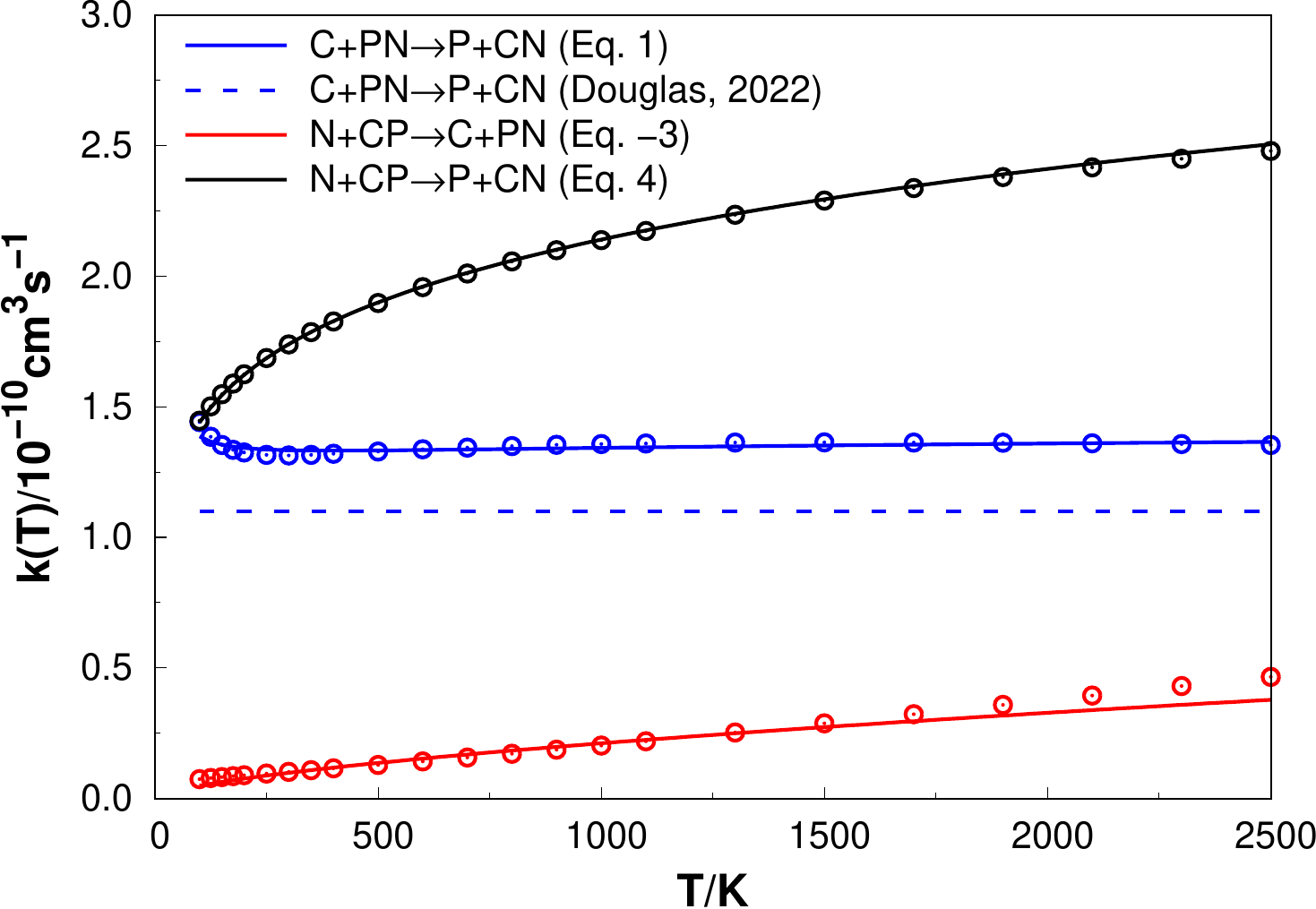}
    \caption{Rate coefficients as a function of temperature obtained for the exoergic reactions. The points refer to the calculated values at fixed temperatures and the curves correspond to a fit to our results using the modified Arrhenius formula. The rate coefficient proposed by \citet{DOU22:99} for the \ce{C + PN -> P + CN} reaction is shown as a dashed line for comparison.}
    \label{fig:rates}
\end{figure}

\begin{figure}
	\includegraphics[width=\columnwidth]{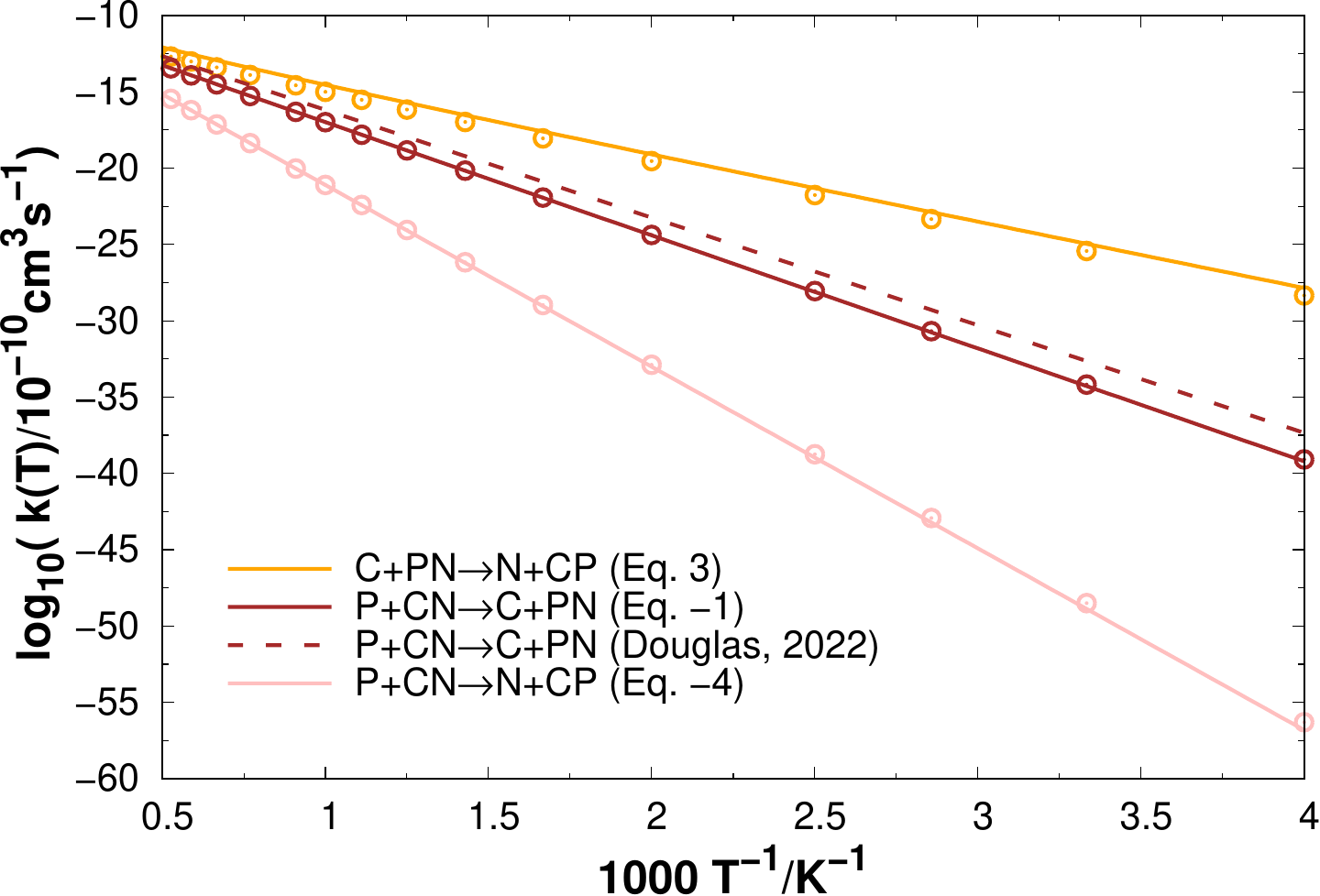}
    \caption{Rate coefficients as a function of inverse temperature obtained for the endoergic reactions. The points refer to the calculated values at fixed temperatures and the curves correspond to a fit to our results using the modified Arrhenius formula. The rate coefficient proposed by \citet{DOU22:99} for the \ce{P + CN -> C + PN} reaction is shown as a dashed line for comparison.}
    \label{fig:rates-endo}
\end{figure}

\subsection{Reaction rates}
\label{sec:rates} 

Among the reactions addressed in this work, only processes \ce{P + CN -> C + PN} and \ce{N + CP -> C + PN}  have rate constants available in the UMIST astrochemical database \citep{UMIST}, both having the same value of $k=3\times 10^{-10}$\,cm$^{3}$\,s$^{-1}$ and none of them are currently present in KIDA (Kinetic Database for Astrochemistry) \citep{KIDA}. However, the former reaction is shown here to be endothermic, and the present value available in UMIST may lead to wrong conclusions on astrochemical models. As discussed in section~\ref{sec:prod}, this was estimated based on the analogous nitrogen reaction (N + CN $\rightarrow$ $\mathrm{N_2}$ + C) assuming similar chemical behaviour between the involved species. However, the PN molecule is not as thermodynamically stable as N$_2$, and we have shown that these chemical processes are actually quite different and present very distinct energy landscapes. The \ce{P + CN -> C + PN} reaction for instance, is significantly endoergic and should not present such a high rate coefficient, nor should it be independent of temperature. As for the \ce{N + CP -> C + PN} reaction, our calculations predict the existence of a small entrance barrier for the most straightforward mechanism, contrary to what occurs for the analogous nitrogen reaction \citep{MOS01:4156}. Therefore, despite being exoergic, one should expect the \ce{N + CP -> C + PN} reaction to present a lower rate constant than the currently assigned one.

We have used our accurate computations at the MRCI-F12/AV(T+d)Z//CAS/AV(T+d)Z + ZPE(CAS/AV(T+d)Z) level on the CPN system to calculate the rate coefficients of reactions in Eqs.~\ref{eq:c+pn1},~\ref{eq:c+pn3},~\ref{eq:n+cp} and their reverses as a function of temperature using the MESS package. The rate coefficients obtained for all six reactions for several different temperature values have been fitted to the modified Arrhenius equation:

\begin{equation} \label{eq:m-arrh} 
    k(T)=\alpha(T/300)^{\beta}\mathrm{exp}(-\gamma/T).     
\end{equation}

\noindent and the fitting parameters ($\alpha$, $\beta$ and $\gamma$ constants) are compiled in Table~\ref{chem-net}. One can also find in this table results from previous studies of our research group as well as the ones proposed by \citet{DOU22:99} and \citet{Garcia2021}. In this way, we offer a wide set of reaction rates for modelling the abundances of P-bearing molecules. We also refer the reader to the supplementary information file where one can find more details about the determination of the rate coefficients of the \ce{PO + N -> PN + O} and \ce{PO + N -> NO + P} reactions.

The calculated rate coefficients with the respective fitted curves for the exoergic reactions are given in Fig.~\ref{fig:rates}. One can notice that these reactions show a very small temperature dependence, as is usually the case for barrierless exoergic processes. As expected, our predictions that the branching ratio between N+CP\ce{->}P+CN and N+CP\ce{->}C+PN largely favours the former, although only the latter is available in UMIST.
For typical ISM conditions, the difference is about two orders of magnitude.

The obtained rate coefficients for the reaction in Eq.~\ref{eq:c+pn1}, which may have an important impact on the abundance of PN in the ISM as the only reaction able to remove this species through neutral-neutral collision, are shown as blue circles in Fig.~\ref{fig:rates}. They are essentially constant over a wide range of temperatures with a mean value of $k=1.35\times 10^{-10}$\,cm$^{3}$\,s$^{-1}$, which is consistent with what is expected from a barrierless exoergic reaction. Nonetheless, a rather unusual increase in the reaction rate coefficients as the temperature decreases can be observed in the low temperature regime ($< 200$ K). This can be explained due to the electronic degeneracy of the carbon atom in the $\mathrm{^3P}$ state when we consider the splitting of the ground energy level due to spin-orbit coupling. This interaction gives rise to three finely separated energy levels with total angular momentum quantum number $J = 0, 1$ and $2$ and electronic degeneracy number $g = 1, 3$ and $5$, respectively. As temperature decreases, the states with lower degeneracy (and also lower energy) become more populated, increasing the values of the rate coefficients due to the higher statistical weight factors associated with these states.

The reactions in Eqs.~\ref{eq:c+pn1} and \,$-$~\ref{eq:c+pn1} have also been approached by \citet{DOU22:99}, who essentially set the rate coefficient of the direct process equal to the capture rate calculated using long-range transition state theory \citep{GEO05:194103}. The rate constant of the reverse process was then calculated by detailed balance. The authors justify this procedure by considering that the reaction in Eq.~\ref{eq:c+pn1} does not have a complex PES and they seem to treat only the CNP intermediate as relevant to those reactions, neglecting other possible multi-step pathways. It should still be a good approximation and, therefore, they are also shown in Figs.~\ref{fig:rates} and~\ref{fig:rates-endo} as dashed lines for comparison. There is clearly good agreement between our results and those obtained by \citet{DOU22:99} for these reactions.

The results for the endoergic reactions are given in Fig.~\ref{fig:rates-endo} in an Arrhenius plot. The activation energies associated with these processes are related to the slope of the curves. As expected, the rate coefficients of these reactions (Eqs. \,$-$~\ref{eq:c+pn1},~\ref{eq:c+pn3} and \,$-$~\ref{eq:n+cp}) present a strong temperature dependence. By analysing the results shown in Fig.~\ref{fig:rates-endo}, it becomes evident that the value assigned in UMIST to the rate constant of the  P + CN $\rightarrow$ C + PN reaction  differs drastically to the calculated ones. According to our calculations, only at $1000$ K the rate coefficient for the P + CN $\rightarrow$ C + PN process reaches $k=1.0\times 10^{-17}$\,cm$^{3}$\,s$^{-1}$, seven orders of magnitude lower than the temperature independent value based on the analogous N reaction \citep{BAU92:411,AGU07:L91}. 
 While high-temperature values are relevant for studies of exoplanetary atmospheres \citep{Alrefaie2022,Fleury2025}, our focus here is on the relatively cold regions of the ISM. The highest rate coefficient in Fig.~\ref{fig:rates-endo} is the one of the reaction in Eq.~\ref{eq:c+pn3}, and it only reaches the order of magnitude of $10^{-10}$\,cm$^{3}$\,s$^{-1}$ at $7000$ K. Hence, we should not expect these endoergic reactions to play a relevant role in the chemical evolution of P-bearing species in cold astronomical environments.

\section{Astrochemical modelling of P-bearing species}
\label{model-results}

\begin{table}
	\caption{Initial elemental abundances used as chemical conditions (see \S~\ref{model-results}).}
	\label{tab-abundances}
    \centering
	\begin{tabular}{lc}
		\hline
		Species & Abundance \\
		\hline
		H$_2$   & 0.5                        \\
		He      & 9.0  $\times$ 10$^{-2}$    \\
		N       & 6.2 $\times$ 10$^{-5}$     \\
		O       & 3.3 $\times$ 10$^{-4}$     \\
		C$^+$   & 1.8 $\times$ 10$^{-4}$     \\
		S$^+$   & 1.5 $\times$ 10$^{-5}$     \\
		Si$^+$  & 1.8 $\times$ 10$^{-6}$     \\
		Fe$^+$  & 2.0  $\times$ 10$^{-7}$    \\
		Na$^+$  & 2.3 $\times$ 10$^{-7}$     \\
		Mg$^+$  & 2.3 $\times$ 10$^{-6}$     \\
		P$^+$   & 7.8 $\times$ 10$^{-8}$     \\
		Cl$^+$  & 3.4 $\times$ 10$^{-8}$     \\
		F       & 1.8 $\times$ 10$^{-8}$     \\
		\hline
	\end{tabular}\\
	\footnotesize{Note: Values were taken from \cite{Jenkins2009,Ruaud2018,Campanha2022}.}
\end{table}

\begin{table*} 
\caption{List of chemical reactions involving phosphorus-bearing species, computed using the Nautilus code, and analysed across the different astrophysical models explored in this study (see~\S~\ref{model-results}).} 
\label{chem-net} 
\begin{tabular}{llllll} 
\hline 
Reaction & \multicolumn{3}{c}{Rate coefficients} & Remark & Method \\ 
                  & $\alpha $ (cm$^3$s$^{-1}$ ) & $\beta$ & $\gamma$ (K)       &               \\ 
\hline 

PN + C $\longrightarrow$ CN + P & $1.29 \times 10^{-10}$ & $0.025$ & $-10.0$ & this work & MESS$^a$ \\ 

P + CN $\longrightarrow$ PN + C & $2.15 \times 10^{-10}$ & $0.118$ & $17000$ & this work & MESS$^a$ \\

N + CP $\longrightarrow$ P + CN & $1.74 \times 10^{-10}$ & $0.172$ & $0$ & this work, new & MESS$^a$ \\

N + CP $\longrightarrow$ C + PN & $9.86 \times 10^{-12}$ & $0.634$ & $0$ & this work, new & MESS$^a$ \\ 


\textcolor{black}{ C + PN  $\longrightarrow$ N + CP } & \textcolor{black}{ $1.09 \times 10^{-11}$ } & \textcolor{black}{ 1.18 } & \textcolor{black}{ 9669 } & this work & MESS$^a$ \\

\textcolor{black}{ P + CN  $\longrightarrow$ N + CP } & \textcolor{black}{$6.01 \times 10^{-10}$ } & \textcolor{black}{ 0.014 } & \textcolor{black}{ 27394 } & this work & MESS$^a$ \\

N + PN $\longrightarrow$ P + N$_2$ & $1.09 \times 10^{-11}$ & $1.02$ & $7919$ & \cite{GOM23:5991} & MESS$^b$ \\ 

\textcolor{black}{ N + PH $\longrightarrow$ P + NH } & \textcolor{black}{$5.62 \times 10^{-13}$ } & \textcolor{black}{ 0.63 } & \textcolor{black}{ 0 } & \cite{GOM23:e011} & VRC-TST$^c$ \\

\textcolor{black}{ P+ NH $\longrightarrow$  N + PH } & \textcolor{black}{$1.93 \times 10^{-13}$ } & \textcolor{black}{ 1.63 } & \textcolor{black}{ 3298 } & \cite{GOM23:e011} & VRC-TST$^c$ \\

P + NH $\longrightarrow$ PN + H & $9.3 \times 10^{-11}$ & $0.18$ & $0.24$ & \cite{GOM23:e011} & VRC-TST$^c$ \\ 

PN + H $\longrightarrow$ P + NH & $6.52 \times 10^{-13}$ & $1.91$ & $23525$ &  \cite{GOM23:e011} & VRC-TST$^c$ \\ 

PH + N $\longrightarrow$ PN + H & $8.8 \times 10^{-11}$ & $0.18$ & $1.01$ & \cite{GOM23:e011} & VRC-TST$^c$ \\ 

PN + H $\longrightarrow$ PH + N & $2.63 \times 10^{-13}$ & $2.25$ & $25333$ &  \cite{GOM23:e011} & VRC-TST$^c$ \\ 

P + O$_2$ $\longrightarrow$ PO + O & $1.44 \times 10^{-12}$ & 1.66  & $704$ & \cite{GOM22:259} & MESS$^b$ \\ 
PO + O $\longrightarrow$ P + O$_2$ & $1.8 \times 10^{-13}$ & $0.79$ & $10054$ & \cite{DOU22:99} & Detailed balance$^d$ \\ 

P + OH $\longrightarrow$ PO + H & $2.28 \times 10^{-10}$ & $0.16$ & $0.37$ &  \cite{Garcia2021} & MESS$^b$ \\ 

PO + H $\longrightarrow$ P + OH & $4.2 \times 10^{-13}$ & $0.82$ & $16791$  & \cite{DOU22:99} & Detailed balance$^d$ \\ 

PH + OH $\longrightarrow$ P + H$_2$O & $1.0 \times 10^{-10}$ & $0.167$ & $0$ & \cite{DOU22:99} & Collision capture rate$^d$ \\ 

P + H$_2$O $\longrightarrow$ PH + OH & $9.0 \times 10^{-10}$ & $0$ & $24319$ & \cite{DOU22:99} & Detailed balance$^d$ \\ 

PH + O $\longrightarrow$ PO + H & $2.0 \times 10^{-10}$ & $0$ & $0$ & \cite{DOU22:99} & Collision capture rate$^d$ \\ 

PO + H $\longrightarrow$ PH + O & $2.8 \times 10^{-10}$ & $0.26$ & $34462$ & \cite{DOU22:99} & Detailed balance$^d$ \\ 

PO + NH $\longrightarrow$ PN + OH & $1.0 \times 10^{-10}$ & $0$ & $0$ & \cite{DOU22:99} & Collision capture rate$^d$ \\ 

PN + OH $\longrightarrow$ PO + NH & $2.6 \times 10^{-10}$ & $0$ & $13002$ &  \cite{DOU22:99} & Detailed balance$^d$ \\ 

PO + Si $\longrightarrow$ P + SiO & $2.1 \times 10^{-10}$ & $0.10$ & $0$ & \cite{DOU22:99} & MESMER$^e$ \\ 

P + H$_2$ $\longrightarrow$ PH + H & $6.7 \times 10^{-10}$ & $0$ & $17349$ & \cite{DOU22:99} & Transition state theory$^d$ \\ 

PH + H $\longrightarrow$ P + H$_2$ & $6.7 \times 10^{-11}$ & $0$ & $346$ & \cite{DOU22:99} & Detailed balance$^d$ \\ 

P + NO $\longrightarrow$ PN + O & $7.1 \times 10^{-11}$ & $0$ & $6427$ & \cite{DOU22:99} & MESMER$^e$ \\ 
PN + O $\longrightarrow$ P + NO & $1.0 \times 10^{-10}$ & $0$ & $3158$ & \cite{DOU22:99} & MESMER$^e$ \\ 
PN + O $\longrightarrow$ PO + N & $2.7 \times 10^{-10}$ & $0$ & $744$ & \cite{DOU22:99} & MESMER$^e$ \\ 
PO + N $\longrightarrow$ P + NO & $2.55 \times 10^{-12}$ & $0$ & $0$ & \citet{MILLAR1987} & Estimation$^f$ \\ 
PO + N $\longrightarrow$ PN + O & $1.53 \times 10^{-11}$ & $-0.31$ & $0$ &  this work$^g$ & MESS$^a$ \\







\hline

\end{tabular}\\ 
    \footnotesize{$^a$ Calculated using the master equation system solver package as described in \S~\ref{sec:coeff}. 
    }\\
    \footnotesize{$^b$ Calculated using the master equation system solver package as described in the associated reference.}\\
    \footnotesize{$^c$ Variable reaction coordinate transition state theory.}\\
    \footnotesize{$^d$ See the associated reference for further details.}\\
    \footnotesize{$^e$ Calculated using the Master Equation Solver for Multi-Energy well Reactions program. 
    See the associated reference for further details.}\\
    \footnotesize{$^f$ See the supplementary information file for discussion.}\\
    \footnotesize{$^g$ Obtained in the present work by employing the master equation system solver package (as described in \S~\ref{sec:coeff}) based on accurate electronic structure calculations carried out in our previous work \citep{SOU21:2} on the PNO system. See the supplementary information file for  a further discussion.}
\end{table*}

Evaluating the behaviour of computed reactions in astrochemical models is essential for understanding their impact on the abundances of phosphorus-bearing species. 
The results are obtained from gas-grain modelling using the Nautilus code, which calculates the initial abundances provided in Table~\ref{tab-abundances} and incorporates the chemical reactions listed in Table~\ref{chem-net} into the KIDA network database (see \S~\ref{sec:gas-grain}).
The primary objective is to analyse the temporal evolution of these reactions and predict the abundances of P, PO, PN, and PH under conditions that approximate the transition from diffuse to dense cloud environments.

\subsection{Diffuse/translucent cloud}

In the life cycle of the ISM, diffuse molecular clouds are low density regions primarily composed of molecular hydrogen (H$_2$) and dust, within which translucent clouds are embedded. These clouds play a crucial role in the early stages of star formation, as they provide the necessary environment for the collapse of matter under gravitational forces, eventually leading to the birth of stars. To investigate the formation of P-bearing species in a diffuse molecular environment, we modelled the system under physical conditions with a gas density of 10$^3$ cm$^{-3}$, a standard gas-to-dust ratio mass ratio of 100:1 \citep{Tricco2017}, and a temperature of 70~K \citep{Corbi2018}. The initial elemental composition of the gas, including both atoms and ions, is presented in Table~\ref{tab-abundances}.

\begin{figure}
    \centering
    \includegraphics[width=1\linewidth]{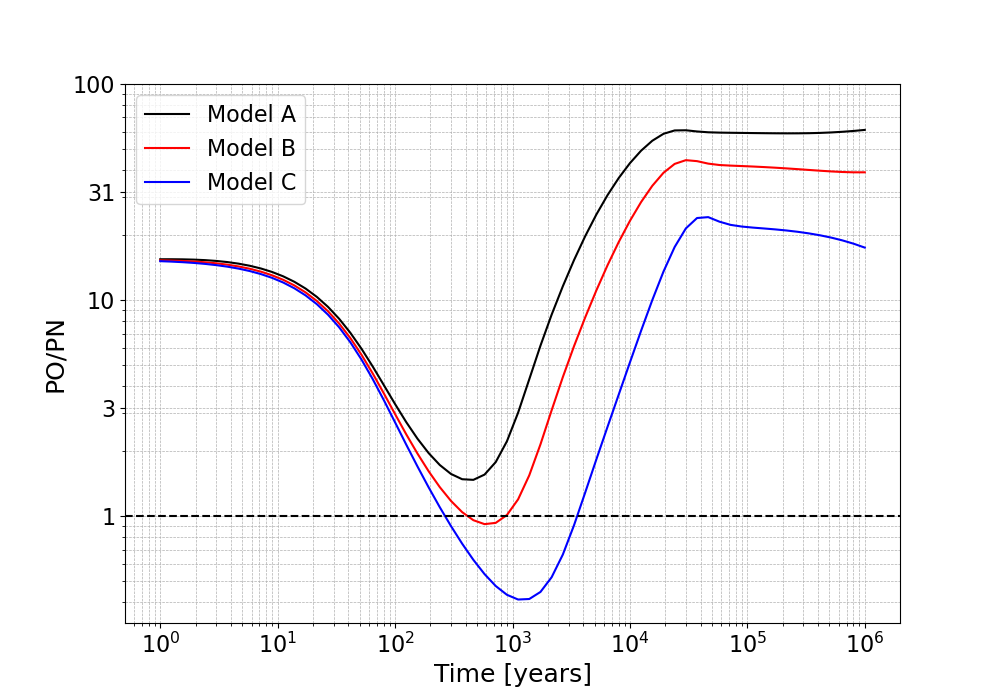}
    \caption{Ratios of PO/PN as a function of time in the diffuse/translucent cloud model: a) $\zeta_{\rm CR}$=1.3 $\times$ 10$^{-16}$ s$^{-1}$ and $\zeta_X$=1 $\times$ 10$^{-17}$ s$^{-1}$; b) $\zeta_{\rm CR}$=6.3 $\times$ 10$^{-17}$ s$^{-1}$ and $\zeta_X$=5 $\times$ 10$^{-18}$ s$^{-1}$; and c) $\zeta_{\rm CR}$=1.3 $\times$ 10$^{-17}$ s$^{-1}$ and $\zeta_X$=1 $\times$ 10$^{-18}$ s$^{-1}$. The dashed line signifies the threshold at which PO/PN equals 1.}
    \label{fig:PO/PN}
\end{figure}

By examining the PO and PN abundances under varying cosmic ray and X-ray ionization rates, we observed corresponding changes in their ratios. The results are exhibited in Fig.~\ref{fig:PO/PN}, which shows the resultant PO/PN ratios as a function of time for the different models: a), b) and c). It is observed that early in the simulation ($t <$ 10$^2$~yr), the impact of the ionization rates on the ratios is less pronounced. However, as time evolves, the differences in PO/PN ratios among the models become  significantly more pronounced. In the final stages ($t >$ 10$^5$~yr), the models show PO/PN  ratios constrained within an interval of roughly 15 to 60 as the cosmic ray ionization rates increase from 1.3 $\times$ 10$^{-17}$ s$^{-1}$ to 1.3 $\times$ 10$^{-16}$ s$^{-1}$. Higher ionization rates and energetic phenomena might affect PO chemistry, influencing the abundance of PO relative to PN. This result is consistent with previous studies that 
 combined quantum chemical and kinetic calculations of PO, also evaluating velocity effects within the context of chemical models in  shocked regions (\citealt{Garcia2021} and references therein).
In comparison to other cases, high PO/PN ratios have been observed in comet 67P/Churyumov–Gerasimenko, \cite{RIV20:1180} estimated that PO is significantly more abundant than PN, with PO/PN ratios exceeding 10.

\begin{figure}
    \centering
    \includegraphics[width=1\linewidth]{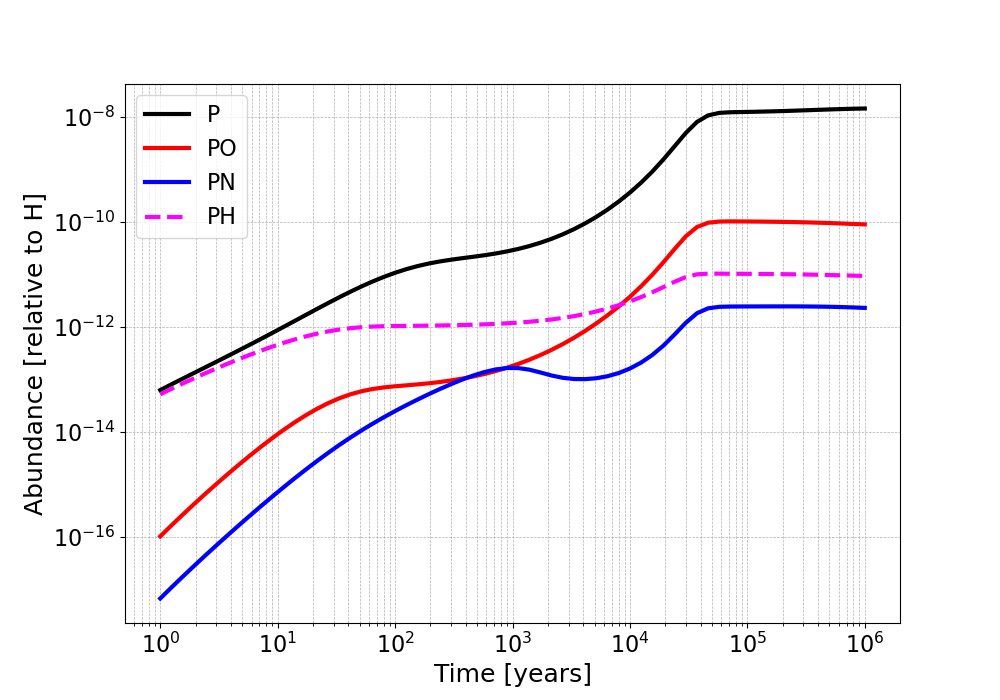}
    \caption{Evolution of the abundances of P, PO, PN and the radical PH (dashed line) in the diffuse/translucent cloud model, adopting a H$_2$ gas density and temperature of 10$^3$ cm$^{-3}$ and 70~K, respectively, with $\zeta_{\rm CR}$=6.3 $\times$ 10$^{-17}$ s$^{-1}$ and $\zeta_X$=5 $\times$ 10$^{-18}$ s$^{-1}$.}
    \label{fig:modelo1}
\end{figure}

We calculated the abundances of phosphorus-bearing species as a function of time, maintaining the gas temperature and density conditions of the diffuse/translucent cloud described previously. To account for the significance of ionization rates, we utilized model B to compute the abundances, since it represents an intermediate case with $\zeta_{\rm CR} = 6.3 \times 10^{-17}$ s$^{-1}$ and $\zeta_X = 5 \times 10^{-18}$ s$^{-1}$. The results are shown in Fig.~\ref{fig:modelo1}, which displays the abundances of P, PO, PN, and the radical PH, over a timescale of 10$^6$~yr. The analysis yields key findings,  such as the elemental phosphorus remains the most abundant species throughout the simulation. Furthermore, PO consistently exhibits higher abundances than PN. Notably, the predicted abundances of P, PO, PN, and PH (phosphinidene) reach a quasi-steady state after approximately 10$^5$ years--significantly earlier than the typical chemical age of 10$^6$ years, adopted for diffuse molecular clouds (e.g. \citealt{Albertsson2014}). In this work, we include the results for PH, represented by a dashed line in Fig.~\ref{fig:modelo1}. Compounds such as phosphinidenes are unstable and highly reactive, presenting a significant challenge for detection. Nevertheless, they play a crucial role in phosphorus chemistry. Specifically, PH contributes to the formation of both PO and PN through neutral reactions: PH + O $\longrightarrow$ PO + H and PH + N $\longrightarrow$ PN + H (Table~\ref{chem-net}).

\subsection{Dense molecular cloud}

A second model of phosphorus chemistry was computed under dense cloud conditions. In this context, the final abundances from the diffuse/translucent model become the initial abundances for the dense cloud. Additionally, considering the radiative processes related to gas cooling, we computed the abundances based on a H$_2$ density of 10$^4$ cm$^{-3}$ and a temperature of 30~K. It is important to note that molecular clouds provide effective shielding against UV and X-ray photons. Consequently, for the dense cloud simulation we adopted $A_V$=10~mag and utilized the lowest ionization rates tested in this study, as discussed in Model C,  $\zeta_{\rm CR}$=1.3 $\times$ 10$^{-17}$ s$^{-1}$ and $\zeta_X$=1 $\times$ 10$^{-18}$ s$^{-1}$. Previous research has highlighted the complexities involved in accurately simulating the transitions from low-density, ionized interstellar medium (ISM) gas to dense, neutral cloud environments (e.g. \citealt{Phan2018}). Thus, we aim to investigate the phosphorus network with a focus on the following objectives: examining the evolution of abundances, identifying key reaction contributions, analysing implications for steady-state conditions, and the dynamic chemical balance over time scales ranging from 1 $\times$ 10$^6$ to 1 $\times$ 10$^8$~yr.

\begin{figure}
    \centering
    \includegraphics[width=1\linewidth]{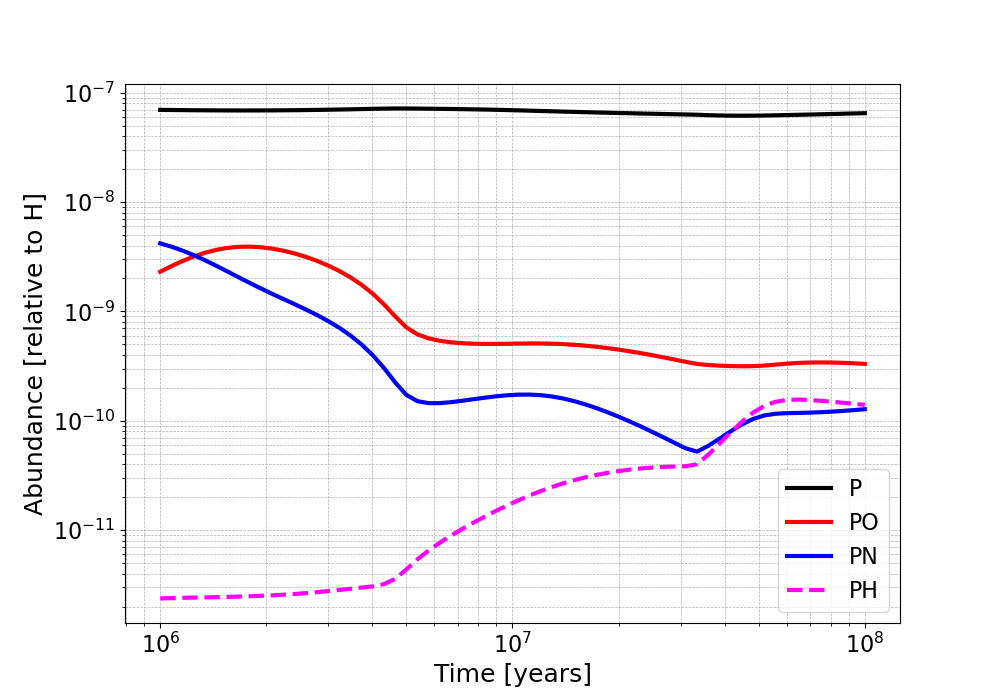}
    \caption{Similar to Fig.~\ref{fig:modelo1}, depicting a  stage from 10$^6$~yr to 10$^8$~yr for a dense cloud model. The parameters are set to a temperature of 30 K, a H$_2$ density of 10$^4$ cm$^{-3}$ and adopting $\zeta_{\rm CR}$=1.3 $\times$ 10$^{-17}$ s$^{-1}$ and $\zeta_X$=1 $\times$ 10$^{-18}$ s$^{-1}$.}
    \label{fig:dense-cloud}
\end{figure}

\begin{figure*}
    \centering
    \includegraphics[width=0.47\linewidth]{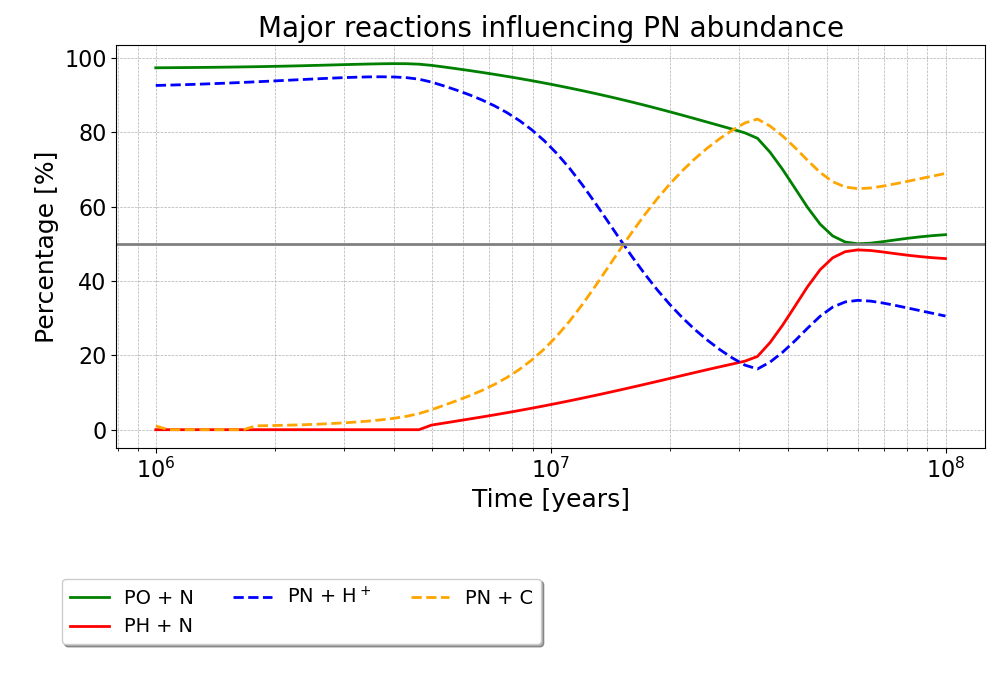}
    \includegraphics[width=0.47\linewidth]{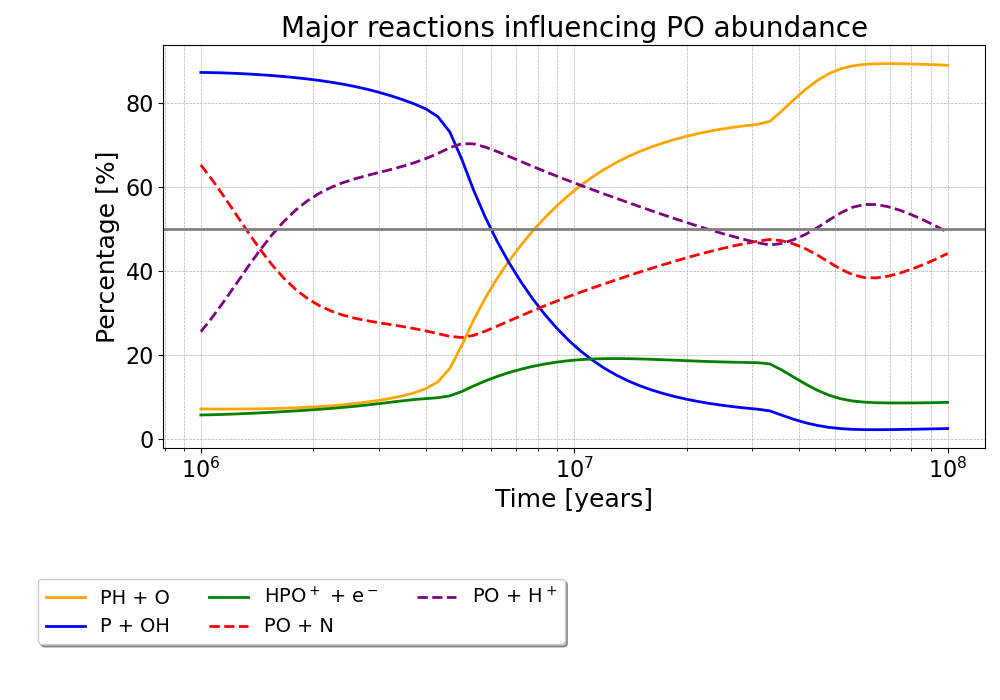}
    \caption{Major reactions for a) PN and b) PO. The contributions of each reaction are represented as percentages over time. Solid and dashed lines differentiate production and destruction mechanisms, respectively. The time is presented on a logarithmic scale to highlight contributions beyond 10$^6$ years, with a horizontal gray line at 50 \% for assessing balance throughout the duration of the long-term trends in the chemical network.}
    \label{fig:sstate-po-pn}
\end{figure*}

The results of this model are presented in Fig.~\ref{fig:dense-cloud}, with the principal contributions of key reactions analysed in Fig.~\ref{fig:sstate-po-pn}. The temporal evolution of the abundances of P, PO, PN, and PH are presented analogously  to Fig.~\ref{fig:modelo1}; however, the analysis is conducted over an extended timescale ranging from 10$^6$ to 10$^8$ years. This hypothetical prolonged period aims to discuss abundance stabilization and the dynamics of the chemical network.  As shown in Fig.~\ref{fig:dense-cloud}, the abundance curves of PO and PN exhibit similar trends, with values fluctuating between $\sim$~5 $\times 10^{-11}$ and 5 $\times 10^{-8}$. Except during the initial phase, the model shows that PO maintains a higher abundance than PN throughout the evolution. Consequently, in the final stage, the PO abundance reaches approximately 3.3 $\times 10^{-10}$, whereas the PN abundance, closely resembling that of PH, stabilizes around 1.3 $\times 10^{-10}$.
Figure~\ref{fig:dense-cloud} also includes the abundance curve of PH, which appears decoupled from the behaviour of PO and PN. The formation of PH is primarily driven by the recombination of molecular ions with electrons, as in reactions like HPO$^+$ + e$^-$ $\longrightarrow$ O + PH and HPN$^+$ + e$^-$ $\longrightarrow$ N + PH. In contrast, its destruction pathways contribute to the synthesis of PO and PN, as seen in the diffuse/translucent model.

We have identified key mechanisms influencing the chemistry of PN and PO. In Fig.~\ref{fig:sstate-po-pn} (a) and (b), we present the percentage contributions of the principal formation and destruction reactions for PN and PO, respectively, as a function of time. In the case of PN (Fig.~\ref{fig:sstate-po-pn} a),  it is important to note that its primary formation pathway is attributed to the PO+N reaction,  which initially contributes slightly less than 100~\% and gradually decreases to approximately 50 \%. Furthermore, the reaction of PH with N contributes to PN formation (e.g., \citealt{GOM23:e011}), though to a lesser degree. Initially, this pathway contributes for only a minor fraction of the total PN production but gradually increases, reaching approximately 50~\% in the final stage of the model. Regarding the destruction mechanisms, PN is primarily destroyed through two pathways involving reactions with H$^+$ and C, with the latter being one of the major reactions explored in this work. The curves associated with these mechanisms exhibit a degree of symmetry, which may arise from the complex interactions inherent in the computations performed by the code. This symmetry indicates a certain predictability and alternation; for instance, at $t\thickapprox 1.5\times 10^7$ yr it can be observed that the contribution of H$^+$ to PN destruction diminishes as the contribution from C becomes more significant.

In the analysis of major reactions influencing PO abundance (Fig.~\ref{fig:sstate-po-pn} b), the primary formation mechanism is the reaction between P and OH (e.g., \citealt{Garcia2021}), which initially accounts for $\sim$~88~\% of the PO production. However, as the chemical network evolves, its contribution gradually decreases toward the final stage of the model. In contrast, the PH+O reaction follows an opposite trends, since it begins with a minimal contribution but steadily increases. As a result, the contribution profiles of these two pathways shift over time, intersecting at approximately $t=6\times 10^6$~yr. To a lesser extent, the reaction HPO$^+$ +e$^-$ also contributes to the PO production, though their impact is significantly lower than that  of the P+OH and PO+H pathways. Regarding the major destruction pathways of PO, two key reactions contribute to its  depletion: PO+N and PO+H$^+$. The relative significance of these mechanisms evolves over time. Initially, the PO+N reaction  is the dominant process; however, its contribution gradually declines, allowing the PO+H$^+$ reaction to become the primary destruction pathway in the later stages.  

Our findings indicate that the computed kinetic equations and percentage contributions reveal a pattern that is both somewhat predictable and alternates in behaviour.  The formation routes (P+OH and PH+O) appear to establish a dynamic equilibrium with the destruction mechanisms (PO+N and PO+H$^+$), suggesting that the steady-state abundance of PO is regulated by the competition between these pathways, which in turn has an effect on the chemistry and abundance of PN. In future work, we will explore this equilibrium in greater detail, assessing its implications for the predictability of chemical abundances under interstellar conditions.

\section{PO/PN ratios and observational implications}

PO and PN have been identified across a range of astronomical environments, including star-forming regions, protostellar objects, shock regions, and giant molecular clouds. Generally, these molecules have been observed in the interstellar medium and within the circumstellar envelopes of evolved stars. Among the two species, PO is widely recognised as being both more abundant and more prevalent than PN. Studies in the literature consistently indicate that PO is typically more abundant than PN by factors of approximately 1.4 to 3, regardless of the specific physical characteristics of the observed sources \citep{MAR23:47}. In modelling a diffuse/translucent cloud, we consistently found that PO is more abundant than PN, with both species displaying a dependence on factors such as visual extinction and influenced by factors like X-ray and cosmic ray ionization rates. Specifically,  these findings suggest that the PO/PN ratios vary over time and increase as the cosmic-ray ionisation rate rises from $\zeta_{\rm CR}$=1.3 $\times$ 10$^{-17}$ to 1.3 $\times$ 10$^{-16}$ s$^{-1}$. However, under conditions of low  ionisation rates, the PO/PN ratio may drop below 1. This fluctuation highlights the sensitivity of phosphorus chemistry to ionisation mechanisms, emphasising its critical role in modelling the relative abundances of PO and PN over time.

In a study of diffuse and translucent clouds, \citet{CHA20:A54} conducted observations along the line of sight towards B0355+508, focusing on the analysis of P-bearing species. For PO and PN, the derived upper abundance limits were determined to be $<$ 5.0 $\times$ 10$^{-10}$ and $<$4.9$\times$10$^{-11}$, respectively.  Based on our theoretical predictions (Fig.~\ref{fig:modelo1}), we estimate PO and PN abundances reaching  $\sim$~1$\times$10$^{-10}$ and 2$\times$10$^{-12}$, respectively, which are consistent with these upper limits. In a study of P-bearing species on a galactic scale, \cite{KOE23:292} determined the abundances of PO and PN in a star-forming region in the outer Galaxy to be $\sim$ 2.0$\times$10$^{-11}$ and 3.0 $\times$10$^{-12}$, respectively. They also estimated a fit for the PO and PN abundances as a function of distance from the Galactic Center, suggesting a slight gradient—a decrease in abundance by factors of approximately 1.5 and 2.3 for PO and PN, respectively, between 8.5 and 22.6~kpc.

In a model simulating the conditions of a denser cloud ($n \simeq 10^4$ cm$^{-3}$), adopting the standard cosmic ray ionization rate  ($\zeta_{\rm CR}$=1.3 $\times$ 10$^{-17}$~s$^{-1}$), we predicted the abundances of P, PO, PN, and PH, along with the major reaction pathways involving these species. In this model, the final abundances of PO and PN  were estimated to be $\sim 3 \times 10^{-10}$ and 1 $\times 10^{-10}$, respectively. Through a detailed examination of the chemical reactions involving both species, we identified that the primary formation pathway of PN occurs via the reaction of PO with N. This finding is further supported by recognizing that the primary mechanism for PO destruction is its reaction with N, which leads to PN formation (see Fig.~\ref{fig:sstate-po-pn}).  In an early work, \cite{THO84:139}, using laboratory results and kinetic modelling, highlighted the significance of PO as a pivotal species in phosphorus chemistry, serving as a primary reservoir of phosphorus with the potential for other species, such as PN and CP. Consequently, PO is a molecule that, once synthesised, accumulates in sufficient quantities to support the formation of additional species, such as PN, without being considerably depleted.

\cite{BER21:55} reported relative abundances of PO $\sim$ 1.6 $\times$ 10$^{-10}$ and PN $\sim$ 6.1 $\times$ 10$^{-11}$ in the Orion Plateau, yielding a PO/PN ratio of approximately 3. They discussed the presence of PO in high-mass star-forming regions, with the potential for its occurrence in low-mass sources as well. In the Orion-KL region, the clear association between PO and PN suggests that these two molecules are likely formed together through interconnected reactions, as it has been found in the present work. In the context of solar-type protostellar objects, the shock region L1157-B1 is characterized by active and complex chemistry.  The detection and formation of various molecules, including HCN, HNC, SiS, HC$_3$N, PO and PN, have been investigated in detail, with a particular focus on shock-induced gas-phase chemistry  \citep{LEF16:3937,Mendoza2018,Mota2021,Lefloch2021}. 
In a recent study, \cite{Lefloch2024} reported that the PO/PN abundance ratios range from 1 to 5.  They argued that the PO/PN ratios observed throughout B1 indicate that the phosphorous chemistry is time-dependent. It has also been suggested that neither species is sputtered from icy mantles; rather, PO forms in the gas phase, followed by the subsequent synthesis of PN (see also \citealt{LEF16:3937}).  \cite{Garcia2021} carried out quantum chemical computations to determine the reaction rates and branching ratios for the reactions P+OH $\rightarrow$ PO+H and P+H$_2$O$\rightarrow$ PO + H$_2$. Additionally, they conducted chemical modelling in a shocked region, predicting PO/PN ratios exceeding 1, ranging from 3 to 9.  \cite{Wurmser2022} conducted a search for PN and PO lines toward a sample of solar-type protostars with well-characterized outflows. They identified a relatively narrow range of PO/PN ratios, extending from 0.6 to 2.2. Their analysis highlighted the critical role of outflows and shocked gas in driving gas-phase phosphorus chemistry in these types of sources. Consistent with findings in the literature, gas-phase phosphorus reactions account for the observed abundances of species such as PO and PN. However, follow-up studies are essential to investigate the chemical dynamics between PO and PN in greater detail, as these species play a fundamental role in complex processes related to prebiotic chemistry and astrobiology.

\section{Conclusions}

Current astrochemical models lack important PN destruction routes \citep{MAR23:47}. Furthermore, reactions such as N + CP $\rightarrow$ PN + C and P + CN $\rightarrow$ PN + C have very uncertain rate constants available in the literature \citep{JIM18:128J}, and the rate coefficient of the nitrogen analogous reaction (N + CN $\rightarrow$ $\mathrm{N_2}$ + C) is often used to describe the chemical evolution of these P-bearing molecules in the interstellar medium \citep{MAR23:47}. In this work we tackled these issues by performing high-accuracy theoretical calculations (MRCI-F12/AV(T+d)Z//CAS/AV(T+d)Z and CCSD(T)-F12/VTZ-F12//M06-2X/AV(T+d)Z) and employing transition state theory to study the potential energy landscape of the CPN system and to elucidate the kinetics of the C($\mathrm{^3P}$) + PN($^1\Sigma^+$) destruction of phosphorus nitride. The outcomes of collisions involving N($\mathrm{^4S}$) + CP($^2\Sigma^+$) and P($\mathrm{^4S}$) + CN($^2\Sigma^+$) species were also investigated. We have then integrated a revised chemical network for P-bearing molecules into the default KIDA network and ran the Nautilus gas-grain Astrochemical code to investigate the temporal evolution of these reactions within diffuse and dense interstellar cloud conditions.

We have found several mechanisms for PN destruction by collisions with a carbon atom that can lead to P($\mathrm{^4S}$) + CN($^2\Sigma^+$) formation without an activation barrier. This is the only product channel that is open for the C($\mathrm{^3P}$) + PN($^1\Sigma^+$) reaction for whatever collision energy. Regarding the kinetics, the temperature-dependent rate coefficients for the reactions addressed in this work were fitted to the modified Arrhenius equation: $k(T)=\alpha(T/300)^{\beta}\mathrm{exp}(-\gamma/T)$. The C($\mathrm{^3P}$) + PN($^1\Sigma^+$) $\rightarrow$ P($\mathrm{^4S}$) + CN($^2\Sigma^+$) reaction has a rate coefficient with parameters $\alpha=1.29\times 10^{-10}$\,cm$^{3}$\,s$^{-1}$, $\beta=0.025$ and $\gamma=-10.0$\,K, consisting in a relevant PN destruction route that should be important to properly model the abundances of P-bearing species in the ISM.

The N($\mathrm{^4S}$) + CP($^2\Sigma^+$) formation processes were found to be considerably endoergic and mechanisms towards CP formation should not take place unless very high temperatures are involved. The destruction of CP, however, is expected to readily occur by collisions with nitrogen atoms even at low temperatures. The obtained rate coefficient for the N($\mathrm{^4S}$) + CP($^2\Sigma^+$) $\rightarrow$ P($\mathrm{^4S}$) + CN($^2\Sigma^+$) reaction has $\alpha=1.74\times 10^{-10}$\,cm$^{3}$\,s$^{-1}$, $\beta=0.172$ and $\gamma=0$ and is higher than the one obtained for the N($\mathrm{^4S}$) + CP($^2\Sigma^+$) $\rightarrow$ C($\mathrm{^3P}$) + PN($^1\Sigma^+$), which has $\alpha=9.86\times 10^{-12}$\,cm$^{3}$\,s$^{-1}$, $\beta=0.634$ and $\gamma=0$. To the best of our knowledge, both rate constants are being reported for the first time. We have also concluded that the energy landscape of the CPN system is very different from that of NCN, and the use of the rate coefficient of the nitrogen analogous reaction is not adequate to describe the evolution of reactions within the CPN system.

We have presented two chemical models to predict the abundances of phosphorus-bearing species, specifically P, PO, PN, and PH, under varying astrophysical conditions. In the diffuse cloud scenario, the influence of ionizing agents on phosphorus chemistry was investigated through PO/PN abundance ratios. Our results indicate that models with higher X-ray and cosmic ray ionization rates strongly favour PO formation,  leading to PO/PN ratios significantly greater than 1. In contrast, models with lower ionization rates can result in slightly higher PN abundances, with PO/PN approaching or slightly below 1  around $t=10^3$~yr. In the final phase of the model ($t \thickapprox 1\times 10^6$~yr), PO/PN range from 15 to 60 depending on the ionization rates. A diffuse cloud model was computed for an intermediate ionization rate scenario, assuming
$\zeta_{\rm CR}$=6.3 $\times$ 10$^{-17}$ s$^{-1}$ and $\zeta_X$=5 $\times$ 10$^{-18}$ s$^{-1}$. As a result, quasi-stable abundances of PO and PN were obtained for $t > 10^5$ yr, with values of $\thickapprox 1 \times 10^{-10}$ and  $\thickapprox 2 \times 10^{-12}$, respectively.

For a subsequent second model adopting dense cloud conditions, the chemical network was computed over an extended  hypothetical timescale ranging from $10^6$ to $10^8$ yr, with $\zeta_{\rm CR}$ = 1.3 $\times$ 10$^{-17}$ s$^{-1}$ and $\zeta_X$ = 1 $\times$ 10$^{-18}$ s$^{-1}$. The results indicate that  PO and PN abundances slightly fluctuate over time, with values ranging from  5 $\times 10^{-11}$ to 5 $\times 10^{-8}$. The balance of the chemical network and its influence on steady-state abundances were analysed by examining the temporal percentage contributions of key reactions.  PO is consistently the most abundant species compared to PN and PH under these conditions. A detailed analysis of the key formation and destruction pathways revealed that PO is primary produced via the  P+OH and PH+O reactions, whose relative contributions vary over time.  With a lower contribution  (up to 20~\%), the HPO$^+$ + e$^-$ reaction also plays a role in the PO formation. Regarding PO destruction mechanisms, the most significant channels are PO+N and PO+H$^+$. An analysis of the major formation and destruction pathways of PN was carried out. The primary formation of PN occurs via the reaction of PO with N, contributing more than 50 \%. This underscores the critical role of PO as a reservoir species that accumulates sufficiently to drive subsequent PN formation.  Nevertheless, PO remains as the most abundant P-bearing molecule.  The second most important mechanism, the PH+N reaction, contributes to the remaining percentage. Finally, regarding PN destruction, we find that its depletion is primarily driven by reactions with H$^+$ and C.  These findings align with earlier studies, confirming the interplay of phosphorus chemistry under varying ionization environments.

In future work, we intend to further enhance the accuracy of the obtained rate coefficients for the approached chemical reactions by performing quasi-classical trajectory simulations. It involves, however, the calculation of a huge set of \textit{ab initio} energies for sufficiently many nuclear arrangements so that an analytic expression can be fit to such data and used to model the interaction potential. Despite being a very time-consuming process, a reliable analytical model offers the advantage of providing fast, continuous and differentiable representations of the PES \citep{ROC20:106913}. A robust scheme, called the CHIPR method, has been proposed by Varandas \citep{VAR13:054120,VAR13:chapter}, and the code of a general program for analytically representing PESs of any triatomic molecules with CHIPR has already been provided \citep{ROC20:106913}. This method should be used as the next step in our study of the CPN system.

\begin{acknowledgements}

We would like to thank the anonymous reviewer for their insightful comments. The authors would like to thank the financial support provided by the Coordenação de Aperfeiçoamento de Pessoal de Nível Superior - Brasil (CAPES, finance code 001), Conselho Nacional de Desenvolvimento Científico e Tecnológico (CNPq, grants 311508-2021-9 and 405524-2021-8), Fundação de Amparo à Pesquisa e Inovação do Espírito Santo (FAPES), Fundação de Amparo à Pesquisa do Estado de Minas Gerais (FAPEMIG, grants APQ-03705-23 and RED-00045-23), Centro Federal de Educação Tecnológica de Minas Gerais (CEFET-MG) and Universidade Federal do Espírito Santo (UFES).  E.M. acknowledges support under the grant "María Zambrano" from
the UHU funded by the Spanish Ministry of Universities and the "European Union NextGenerationEU." This project has also received funding from the European Union's Horizon 2020 research and innovation program under Marie Sklodowska-Curie grant agreement No. 872081, and grant PID2022-136228NB-C21 (M.C.) funded by MCIN/AEI/10.13039/501100011033, and, as appropriate, by "ERDF A way of making Europe", the "European Union", or the "European Union NextGenerationEU/PRTR". This work is also supported by the Consejería de Transformación Económica, Industria, Conocimiento y Universidades, Junta de Andalucía and European Regional Development Fund (ERDF 2021-2027) under the project EPIT1462023 (M.C. \& E.M.).

\end{acknowledgements}


\bibliographystyle{aa} 
\bibliography{biblio.bib} 

\end{document}